\documentclass{article}
\usepackage{ijcai17}
\usepackage{graphicx}
\usepackage{times}
\usepackage{multirow}
\usepackage{comment}
\usepackage{booktabs}
\usepackage{amssymb}
\usepackage{amsfonts}
\usepackage{latexsym}
\usepackage{amsmath}
\usepackage{graphicx}
\usepackage{subfigure}
\usepackage{url}
\usepackage{flushend}
\usepackage{color, colortbl}
\usepackage{color, soul}
\usepackage[table]{xcolor}
\usepackage{multirow}
\usepackage{hyperref}
\usepackage{balance}
\usepackage{fixltx2e}
 
\usepackage[small]{caption}

\title{Network Vector: Distributed Representations of Networks with Global Context}
\author{Hao Wu \\ USC ISI \\ hwu732@usc.edu \And Kristina Lerman \\ USC ISI \\ lerman@isi.edu}

\begin{document}

\maketitle
\begin{abstract}
We propose a neural embedding algorithm called Network Vector, which learns distributed representations of nodes and the entire networks simultaneously. By embedding networks in a low-dimensional space, the algorithm allows us to compare networks in terms of structural similarity and to solve outstanding predictive problems. Unlike alternative approaches that focus on node level features, we learn a continuous global vector that captures each node's global context by maximizing the predictive likelihood of random walk paths in the network. Our algorithm is scalable to real world graphs with many nodes. We evaluate our algorithm on datasets from diverse domains, and compare it with state-of-the-art techniques in node classification, role discovery and concept analogy tasks. The empirical results show the effectiveness and the efficiency of our algorithm. 

\end{abstract}

\section{Introduction}
Applications in network analysis, including network pattern recognition, classification, role discovery, and anomaly detection, among others, critically depend on the ability to measure similarity between networks or between individual nodes within networks.
For example, given an email communications network within an enterprise, one may want to classify individuals according to their functional roles, or given one individual, find another one playing a similar role.

Computing network similarity requires going beyond comparing networks at a node level to measuring their structural similarity. To characterize network structure, traditional approaches extract features such as node degrees, clustering coefficients, eigenvalues, the lengths of shortest paths and so on~\cite{berlingerio2012netsimile,henderson2012rolx,gilpin2013guided}. However, these hand-crafted features are usually heterogeneous and it is often not clear how to integrate them within a learning framework. In addition, some graph features, such as eigenvalues, are computationally expensive and do not scale well in tasks involving large networks. Recent advances in distributed representation of nodes~\cite{perozzi2014deepwalk,tang2015line,grover2016node2vec} in networks created an alternate framework for unsupervised feature learning of nodes in networks. These methods are based on the idea of preserving local neighborhoods of nodes with neural network embeddings. An objective is defined on the proximity between nodes in exploring the network neighborhood with various strategies, mainly Depth-First Search (DFS) and Breadth-First Search (BFS). The objective is optimized using single layer neural network for efficient training. However, the embeddings used for feature representation limit scope to the local context of nodes, without directly exploiting the global context of the network. 
To represent the whole network, these approaches require us to integrate the representations of all nodes, for example, by averaging their representations. However, not all nodes contribute equally to the global representation of the network, and in order to account for their varying importance, aggregation schemes need to weigh nodes, which adds an extra layer of complexity to the learning task.

To address above-mentioned challenge we describe a neural network algorithm called Network Vector, which learns distributed representations of networks that account for their global context. The algorithm is 
scalable to real world networks with large numbers of nodes and can be applied to generic networks such as social networks, knowledge graphs, and citation networks. Networks are compressed into real-valued vectors that preserve the network structure, so that the learned distributed representations can be used to effectively measure network similarity. Specifically, given two networks, even those with different size and topology, the distance between the learned vector representations can be used to measure their structural similarity. In addition, this approach allows us to compare individual nodes by looking at the similarity of their ego-networks, i.e., networks that contain the focal node and all their neighbors and connections between them. 

Our approach is inspired by Paragraph Vector~~\cite{le2014distributed} that learns distributed representations of texts of variable length such as sentences and documents~\cite{le2014distributed}. By exchanging the notions of ordered ``word" sequences in sentences and ``nodes" in paths along edges on networks. 
We learn network representations in a similar way of learning representations of sentences and documents. Specifically, we sample sequences of nodes from a network using random walks, same as in~\cite{perozzi2014deepwalk,grover2016node2vec}. In contrast to existing approaches, the likelihood of next node in a random walk sequence predicted by our algorithm depends not only on the previous nodes, but also on the global context of the network. The global context vector representation of the network is learned to maximize the average predicted likelihood of nodes in random walk sequences sampled from the network. The learned representations can be used as the signatures of the networks for comparison, or as features for classification and other predictive tasks.

We evaluate the algorithm on several real world datasets from a diversity of domains, including citation network of knowledge concepts in Wikipedia, email interaction network, legal citations network, social network of bloggers, protein-protein interaction network and language network. We focus on predictive tasks including role discovery in networks that aims to 
identify individual nodes serving similar roles, inference of analogous relations between concept pairs in Wikipedia and multi-label node classification. 
We compare Network Vector with state-of-the-art feature learning algorithm node2vec~\cite{grover2016node2vec}, LINE~\cite{tang2015line}, DeepWalk~\cite{perozzi2014deepwalk} and feature-based baselines such as node degrees, clustering coefficients and eigenvalues. Experiments demonstrate the superior performance of Network Vector, due to its capacity of learning the global context of the network.

In summary, our contributions are summarized as follows:
\begin{enumerate}
\item We propose Network Vector algorithm, a distributed feature learning algorithm for representing an entire network and its nodes simultaneously. We define an objective function that preserves the local neighborhood of nodes and the global context of the entire network.
\item We evaluate Network Vector on role discovery, concept analogy and node multi-label classification tasks. Experiments on several benchmark datasets from diverse domains show its effectiveness and efficiency.
\end{enumerate}


\section{Related Work}
Our algorithm builds its foundation on learning distributed representations of concepts~\cite{hinton1986learning} . Distributed representations encode structural relationships between concepts and are typically learned using back-propagation through neural networks. Recent advances in natural language processing have successfully adopted distributed representation learning and introduced a family of neural language models~\cite{bengio2003neural,mnih2007three,mikolov2010recurrent,mikolov2013efficient,mikolov2013distributed} to model word sequences in sentences and documents. These approaches embed words such that words in similar contexts tend to have similar representations in latent space. 

By exchanging the notions of nodes in a network and words in a document, recent research~\cite{perozzi2014deepwalk,tang2015line,cao2015grarep,grover2016node2vec} attempt to learn node representations in a network in a similar way of learning word embeddings in neural language models. Our work follows this line of approaches in which nodes in a neighborhood will have similar embeddings in vector space. Different node sampling strategies are explored for characterizing the neighborhood structure. For example, DeepWalk~\cite{perozzi2014deepwalk} samples node sequences from a network using a stream of short first-order random walks, and model them just like word sequences in documents using neural embeddings. LINE~\cite{tang2015line} samples nodes in pairwise manner and model the first-order and second-order proximity between them. GrapRep~\cite{cao2015grarep} extends LINE to exploit structural information beyond second-order proximity. To offer a flexible node sampling scheme, node2vec~\cite{grover2016node2vec} utilizes second-order random walks, and combines Depth-First Search (DFS) and Breadth-First Search (BFS) strategies to explore the local neighborhood structure.

However, existing approaches only consider the local network structures (i.e., the neighborhoods of nodes) in learning node embeddings, but exploit little information of the global structure of the network. Although recent approach GrapRep~\cite{cao2015grarep} attempts to capture long distance relationship between two different nodes, it limits scope to a fixed number of hops. More importantly, existing approaches focus on node representations, and it requires additional effort to compute the representation of the entire network. The simple scheme of averaging the representations of all nodes to represent the network is by no means a good choice as it ignores the statistics of node frequency and their roles in the network. In contrast, we introduce a notion called the \textit{global network vector}, which aims to represent the structural properties of an entire network. 
The global vector representation of the network acts as a memory which is asked to contribute to the prediction of a node accompanying with the node's neighbors, and updated to maximize the predictive likelihood. As a result, our algorithm can simultaneously learn the global representation of a network and the representations of nodes in the network. This is inspired by Paragraph Vector~\cite{le2014distributed}, which learns a continuous vector to represent a piece of text with variable-length, such as sentences, paragraphs and documents.

\section{Network Vector}
We consider the problem of embedding nodes of a network and the entire network into a low-dimensional vector space.
Let $G = \{V, E\}$ denote a graph,  and $V$ is the set of vertices and $E = V \times V$ is the set of edges with weights $W$. The goal of our approach is to map the entire graph to a low-dimensional vector, represented by ${\bf{v}}_G \in \mathbb{R}^d$, and map each node $i$ to a unique vector ${\bf{v}}_i \in \mathbb{R}^d$ in the same vector space. Although the dimensionality of network representation ${\bf{v}}_G$ can be different from that of node representations ${\bf{v}}_i$ in theory, we adopt the same dimensionality $d$ for the ease of computation in real world applications. Suppose that there are $M$ graphs given (e.g., ego-networks of $M$ persons of interest in a social network) and $N$ distinct nodes in the corpus, then there are $(M + N) \times d$ parameters to be learned.

\subsection{A Neural Architecture}
Our approach of modeling networks 
is motivated by 
learning distributed representations of variable-length texts, e.g., sentences and documents~\cite{le2014distributed}. 
The concept of ``words" in a document~\cite{le2014distributed} is replaced by ``nodes" in a network in our modeling. The goal is to predict a node given other nodes in its local context as well as the global context of the network. Text has a linear property that the local context of a word can be naturally defined by surrounding words in ordered sequences. However, networks are not linear. In order to characterize the local context of a node, without loss of generality, we sample node sequences from the given network with second-order random walks in~\cite{grover2016node2vec}, which offer a flexible notion of a node's
local neighborhood by combining  Depth-First Search (DFS) and Breadth-First Search (BFS) strategies. Our learning framework can easily adopt higher-order random walks, but with higher computation cost. Each random walk starts from an arbitrary root node and generates an ordered sequence of nodes with second-order Markov chains. Specifically, consider node $v_a$ that has been visited in the previous step, and the random walk currently reaches node $v_b$.  Consecutively, the next node $v_c$ will be sampled in random walks, with probability: 
\begin{equation}\label{eq:transition_prob}
P(v_c | v_a, v_b) = \frac{1}{Z} M_{bc}^a W_{bc}
\end{equation}
where $M_{bc}^a$ is the unweighted transition probability of moving from node $v_b$ to $v_c$ given $v_a$, $W_{bc}$ is the weight of edge $(v_b, v_c)$, and $Z$ is the normalization term. We define $M_{bc}^a$ as: 
\begin{equation}
  M_{bc}^a  = \begin{cases}
               \frac{1}{p} &\text{if}\ d(v_a, v_c) = 0\\
               1  &\text{if}\ d(v_a, v_c)=1\\
                \frac{1}{q} &\text{if}\ d(v_a, v_c)=2
            \end{cases}
\end{equation}
where $d(v_a, v_c)$ is the shortest path distance between $v_a$ and $v_c$. The parameters $p$ and $q$ control how the random walk biases toward visited nodes in previous step and nodes that are further away. The random walk terminates when $l$ vertices are sampled, and the procedure repeats $r$ times for each root node.

\begin{figure}
\centering	
{\includegraphics[width=0.48\textwidth]{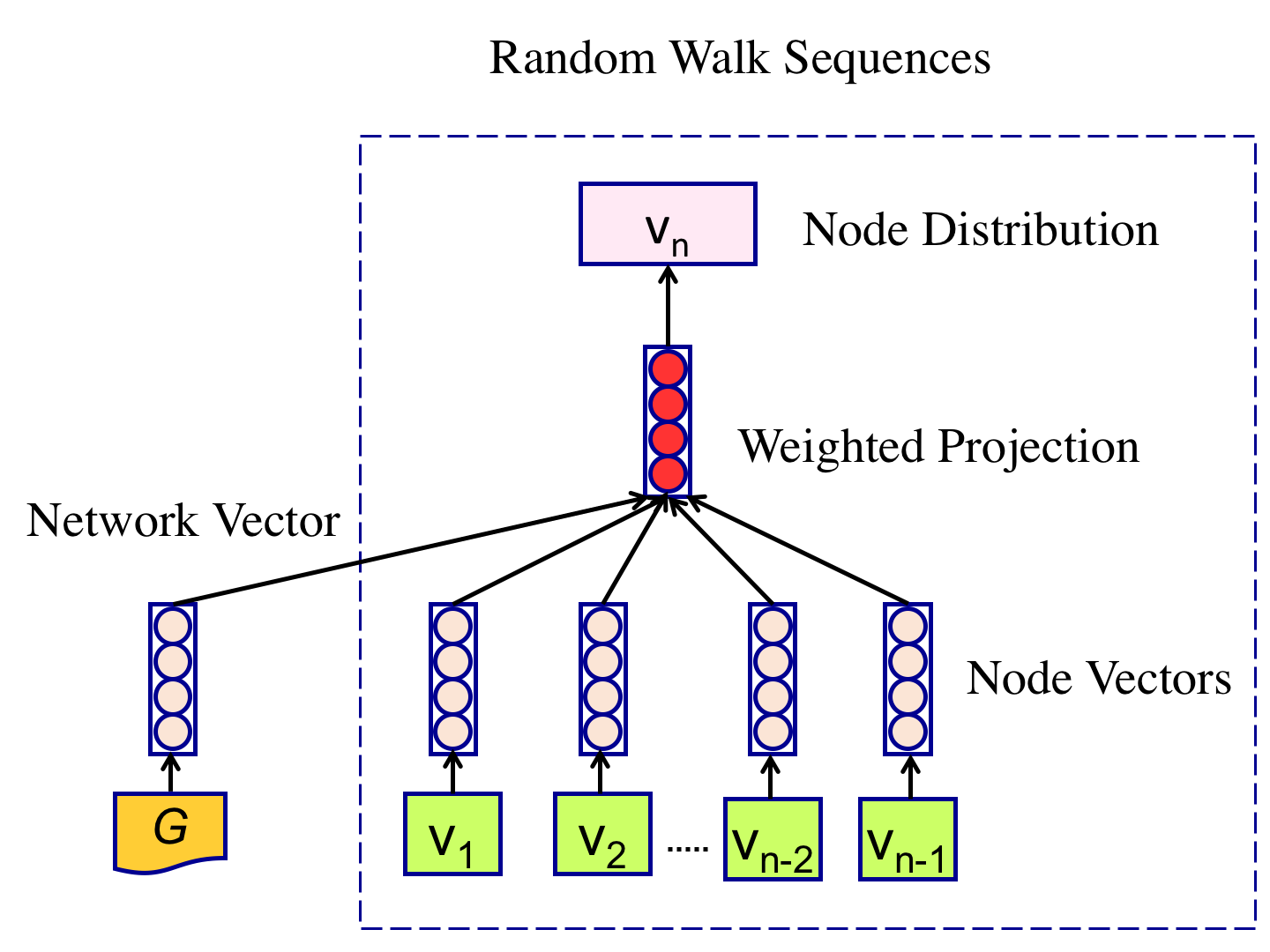}}
\caption{A neural network architecture of Network Vector. It learns the global network representation as a distributed memory evolving with the sliding window of node sequences. \label{fig:networkvector}}
\end{figure}

Figure~\ref{fig:networkvector} illustrates a neural network architecture for learning the global network vector and node vectors simultaneously.
A sliding window  $(v_1, \cdots, v_n)$ with fixed-length $n$ is repeatedly sampled over node sequences. The algorithm predicts the target node $v_n$ given preceding nodes $v_{1:n-1}$ as local context and the entire network $G$ as the global context, with probability $P(v_n | v_{1:n-1}, G)$. Formally, the probability distribution of a target node is defined as:
\begin{equation}\label{eq:objective}
P( v_n | v_{1:n-1}; G)  =  \frac {1} {Z_c} \exp [ -E( G, v_{1:n-1}; v_n) ]
\end{equation}
where ${Z_c} = \sum_{v_m \in V} \exp [ -E( G, v_{1:n-1}; v_m) ] $ is the normalization term. We extend the scalable version of Log-Bilinear model~\cite{mnih2007three}, called vector Log-Bilinear model (vLBL)~\cite{mnih2013learning}. In our model, the energy function $E(G, v_{1:n-1}; v_n)$ is specified as:
\begin{equation}
E( G, v_{1:n-1}; v_n) =  -\hat{\mathbf{v}}^\top\mathbf{v}_{n}
\end{equation}
where $\hat{\mathbf{v}}$ is the predicted representation of the target node:
\begin{equation}
\hat{\mathbf{v}} =  {\bf{v}}_G + \left(\sum_{i=1}^{n-1} \mathbf{c}_i \odot \mathbf{v}_i \right)
\end{equation}
Here $\odot$ denotes the Hadamard (element-wise) product, and $\mathbf{c}_i$ is the weight vector for the context node in position $i$. $\mathbf{c}_i$ parameterizes the context nodes at different hops away from the target node in random walks. The global network vector ${\bf{v}}_G$ is shared across all sliding windows of node sequences. After being trained, the global network vector preserves the structural information of the network, and can be used as feature input for the network. In our model, in order to impose symmetry in feature space of nodes, and activate more interactions between the feature vector ${\bf{v}}_G$ and the node vectors, we use the same set of feature vectors for both the target nodes and the context nodes. This is different from~\cite{mnih2013learning}, where two separated sets of representations are used for the target node and the context nodes respectively. In practice, we find our approach improves the performance of Network Vector.

\subsection{Learning with Negative Sampling}
The global network vector $\mathbf{v}_G$, the node vectors $\mathbf{v}_i$ and the position-dependent context parameters $\mathbf{c}_i$ are initialized with random values, and optimized by maximizing the objective in Eq. (\ref{eq:objective}). Stochastic gradient ascent is performed to update the set of parameters $\boldsymbol \theta = \{\mathbf{v}_G, \mathbf{v}_i, \mathbf{c}_i\}$:
\begin{equation}\label{eq:gradient}
\begin{split}
\Delta \boldsymbol{\theta} &= \epsilon {\nabla_{\boldsymbol{\theta}}} \log P(v_n|v_{1:n-1}, G) \\
                                    & = \epsilon \frac{\partial}{\partial \boldsymbol \theta} \left[\frac{\exp(\hat{\mathbf{v}}^\top\mathbf{v}_{n})} {\sum_{v_m=1}^N \exp(\hat{\mathbf{v}}^\top\mathbf{v}_{m})}\right]
\end{split}
\end{equation}
where $\epsilon$ is the learning rate. The computation involves the normalization term and is proportional to the number of distinct nodes $N$. The complexity of computation is expensive and impractical in real applications.
In our approach, we adopt negative sampling~\cite{mikolov2013distributed} for optimization. Negative sampling represents a simplified version of noise contrastive estimation~\cite{mnih2012fast}, and trains a logistic regression to distinguish between data samples of ${v}_n$ from ``noise" distribution. Our objective is to maximize
\begin{equation}\label{eq:ng}
\log \sigma (\hat{\mathbf{v}}^\top\mathbf{v}_{n}) + \sum_{m=1}^k \mathbb{E}_{v_m} \sim P_n (v) \left[\log \sigma (-\hat{\mathbf{v}}^\top\mathbf{v}_{m}) \right]
\end{equation}
where $\sigma (x) = 1 / (1 + \exp(-x)) $ is the sigmoid function. $P_n (v) $ is the global unigram distribution of the training data acting as the noise distribution where we draw $k$ negative samples of nodes. Negative sampling allows us to train our model efficiently that no longer requires explicitly normalized in Eq. (\ref{eq:gradient}), and hence are more scalable.

\subsection{An Inverse Architecture}
The architecture in Figure~\ref{fig:networkvector} utilizes the linear combination of the global network vector and the context node vectors to predict the target node in a sliding window. Another way of training the global network vector is to model the likelihood of observing a sampled node $v_t$ from the sliding window conditioned on the feature vector ${\bf{v}}_G$, given by
\begin{equation}\label{eq:inverse_probability}
P( v_t | G) =  \frac {1} {Z_G} \exp [ -E( G; v_t) ]
\end{equation}
where $Z_G$ is the normalization term specific to the feature representation of $G$. The energy function $E( G; v_t)$ is:
 \begin{equation}
E( G; v_t) =  -{\mathbf{v}}_G^\top\mathbf{v}_{t}
\end{equation}
This architecture is a counterpart of the Distributed Bag-of-Words version of Paragraph Vector~\cite{le2014distributed}. However, this architecture ignores the order of the nodes in the sliding window and perform poorly in practice when it is used alone. We extend the framework by simultaneously training network and node vectors using a Skip-gram~\cite{mikolov2013efficient,mikolov2013distributed} like model. The model additionally maximizes the likelihood of observing the local context $v_{{t-n}:{t+n}}$ (excluding $v_t$) for the target node $v_t$, conditioned on the feature representation of $v_t$. Unfortunately, modeling the joint distribution of a set of context nodes is not tractable. This problem can be relaxed by assuming the node in different context positions are conditionally independent given the target word:
\begin{equation}
P(v_{{t-n}:{t+n}} | v_t) =  \prod_{i=t-n}^{t+n}  P(v_i | v_t)
\end{equation}
where $P(v_i | v_t)  = \frac {1} {Z_t} \exp [ -E(v_t; v_i) ]$. The energy function is:
\begin{equation}
E(v_t; v_i) =  \mathbf{v}_t^\top (\mathbf{c}_i \odot  \mathbf{v}_i)
\end{equation}
The objective is to maximize the log-likelihood of the product of the probabilities, $P( v_t | G) $ and $P(v_{{t-n}:{t+n}} | v_t)$

\subsection{Complexity Analysis}
The computation of Network Vector consists of two key parts: sampling of node sequences with random walks and optimization of vectors. 
For each node sequence of fixed length $l$, we start from a randomly chosen root node. At each step, the walk visits a new node based on the transition probabilities $P(v_c | v_a, v_b) $ in Eq.~(\ref{eq:transition_prob}). The transition probabilities $P(v_c | v_a, v_b) $ can be precomputed and stored in memory using $O(|E|^2/|V|)$ space. Sampling a new node in the walk can be efficiently done in $O(1)$ time using alias sampling~\cite{walker1977efficient}. The overall time complexity is $O(r|V|l)$ for repeating $r$ times of random walks of fixed length $l$ by taking each node as root. 

The time complexity of optimization with negative sampling in Eq.~(\ref{eq:ng}) is proportional to the dimensionality of vectors $d$, the length of context window $n$ and the number of negative samples $k$. 
It takes $O(dnk)$ time for nodes within the sliding window $(v_1, \cdots, v_n)$. The introduced global vector ${\bf{v}}_G$ requires $O(dk)$ time to optimize, same as any other node vectors in the sliding window. Given $r$ random walks of fixed length $l$ starting from every node, the overall time complexity is $O(dnkr|V|l)$.  To store the node vectors and the global network vector, it requires $O(d|V|+d)$ space.

\subsection{The Property of Network Vector}
The property of the global network vector ${\bf{v}}_G$ in the architecture (as shown in Figure~\ref{fig:networkvector}) can be explained by looking at the objective in Eq. (\ref{eq:objective}). ${\bf{v}}_G$ is part of the input to the neural network, and can be viewed as a term that helps to represent the distribution of the target node $v_n$. The relevant part ${\bf{v}}_G^\top{\bf{v}}_n$ is related logarithmically to the probability $P( v_n | v_{1:n-1}; G)$. Therefore, the more frequently a particular $v_n$ is observed in the data, the larger the value ${\bf{v}}_G^\top{\bf{v}}_n$ will have, and hence ${\bf{v}}_G$ will be closer to ${\bf{v}}_n$ in vector space. The training objective is to maximize the logarithm of the product of all probabilities $P( v_n | v_{1:n-1}; G)$, and the value is related to ${\bf{v}}_G^\top{\bar{\bf{v}}}_n$, where ${\bar{\bf{v}}}_n$ is the expected vector that can be obtained by averaging all observed ${\bf{v}}_n$ in the data. It is also true for Eq.~(\ref{eq:inverse_probability}) in the inverse architecture where the global network vector ${\bf{v}}_G$ is the only input to the neural network, in order to predict every node $v_t$. 

\begin{figure}
\centering	
{\includegraphics[width=0.48\textwidth]{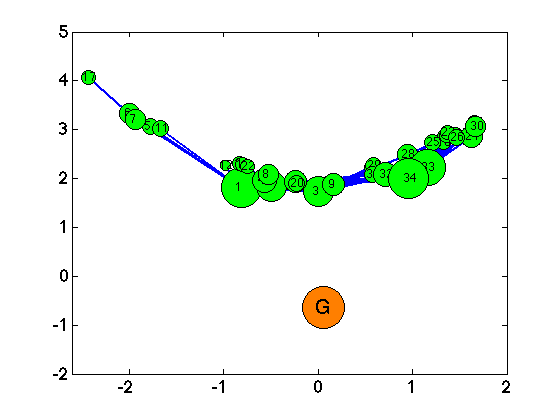}}
\caption{Network Vector is trained on Zachary's Karate network to learn two-dimensional embeddings for representing nodes (green circles) and the entire graph (orange circle). The size of each node is proportional to its degree. Note the global representation of the graph is close to these high-degree nodes. 
 \label{fig:karateexample}}
\end{figure}

\subsubsection{Karate Network}
As an illustrative example, we apply Network Vector to the classic Karate network~\cite{zachary1977information}. The nodes in the network represent members in a karate club, and the edges are social links between the members outside the club. There are 34 nodes and 78 undirected edges in total. We use the inverse architecture to train the vectors. Figure~\ref{fig:karateexample} shows the output of our method in two dimensional space. We use green circles to denote nodes and orange circle to denote the entire graph. The size of a node is proportional to its degree in the graph. We can see that the learned global network vector is close to these high-degree nodes, such as node 1 and 34, which serve as the hubs of two splits of the club. The resulting global vector mostly represent the backbone nodes (e.g., hubs) in the network and compensates the lack of global information in local neighborhoods. 
\begin{table*} [!ht]
\caption{A subset of most cited legal decisions with two distinct patterns of ego-networks. 2-D embeddings of the legal decisions are learned using Network Vector with their ego-networks as input, and mapped to the right figure. Citations of the top 6 cases in the table have a few giant hubs (red dots in the figure), while that of the bottom 6 cases are well connected (blue dots in the figure). \label{table:legal}}
\begin{center}
{\footnotesize
\begin{tabular}{|c|c|c|c|c}
\cline{1-4}
\textbf{ID} &\textbf{Title (with url)}  &\textbf{Page} &\textbf{Year} &\multirow{11}{*}{\includegraphics[scale=0.25]{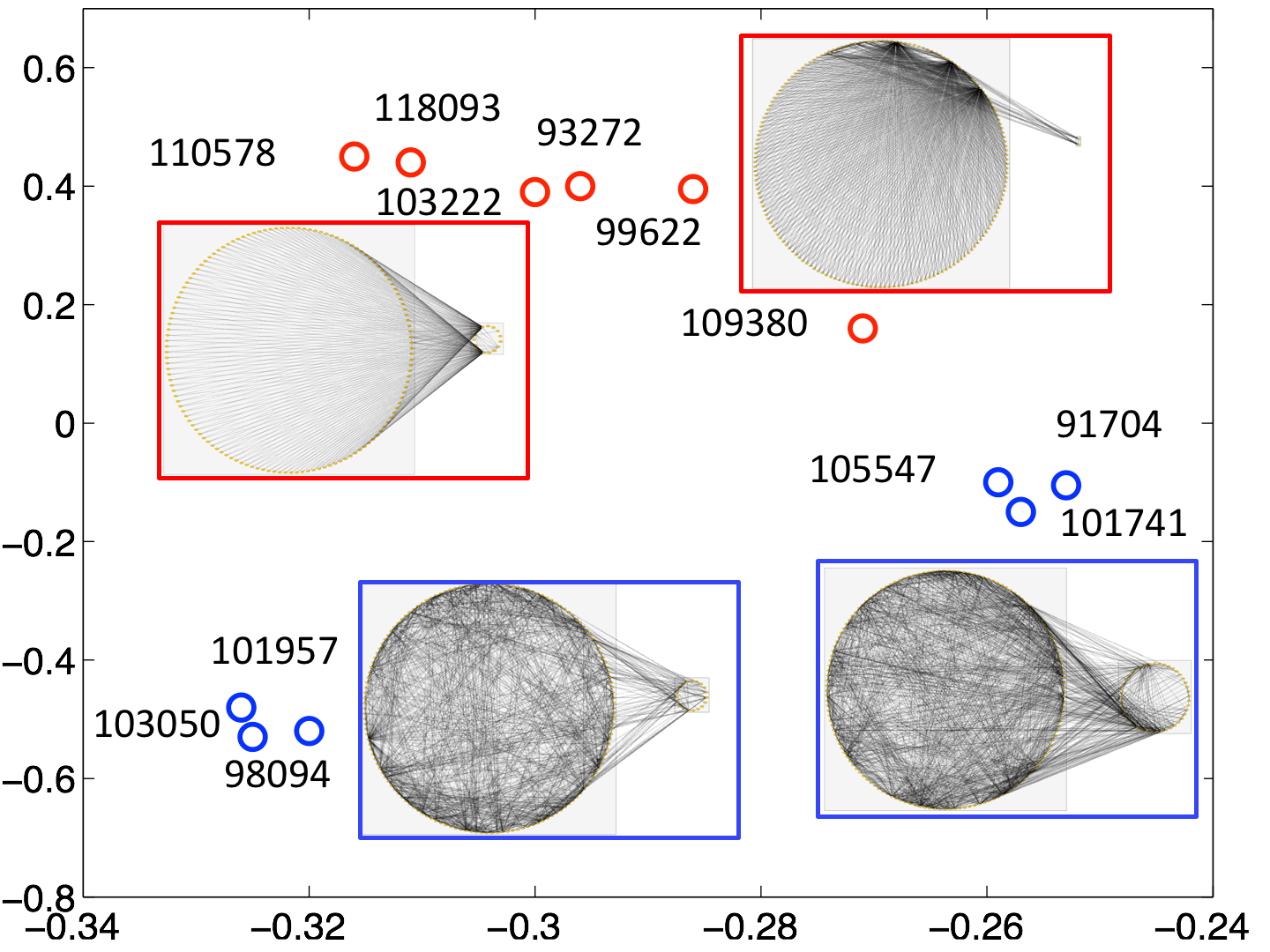}} \\
\cline{1-4}
93272 &{\color{red} \href{https://supreme.justia.com/cases/federal/us/143/339/}{Chicago \& Grand Trunk Ry. Co. v. Wellman} }&143 U.S. 339 &1892 &\\
\cline{1-4}
99622 &{\color{red} \href{https://supreme.justia.com/cases/federal/us/253/412/}{F. S. Royster Guano Co. v. Virginia} } &253 U.S. 412 &1920 &\\
\cline{1-4}
103222 &{\color{red} \href{https://supreme.justia.com/cases/federal/us/307/433/}{Coleman v. Miller}} &307 U.S. 433  &1939 &\\
\cline{1-4}
109380 &{\color{red} \href{https://supreme.justia.com/cases/federal/us/424/1/}{Buckley v. Valeo}} &424 U.S. 1 &1976 &\\
\cline{1-4}
118093 &{\color{red} \href{https://supreme.justia.com/cases/federal/us/520/43/}{Arizonans for Official English v. Arizona}} &520 U.S. 43 &1997 &\\
\cline{1-4}
110578 &{\color{red} \href{https://supreme.justia.com/cases/federal/us/454/46/}{Ridgway v. Ridgway}} &454 U.S. 46 &1981 &\\
\cline{1-4}
91704 &{\color{blue} \href{https://supreme.justia.com/cases/federal/us/118/356/}{Yick Wo v. Hopkins}} &118 U.S. 356 &1886 &\\
\cline{1-4}
98094 &{\color{blue} \href{https://supreme.justia.com/cases/federal/us/232/383/}{Weeks v. United States}} &232 U.S. 383 &1914 &\\
\cline{1-4}
101741 &{\color{blue} \href{https://supreme.justia.com/cases/federal/us/283/359/}{Stromberg v. California}} & 283 U.S. 359 &1931 &\\
\cline{1-4}
101957 &{\color{blue} \href{https://supreme.justia.com/cases/federal/us/287/45/}{Powell v. Alabama}} &287 U.S. 45 &1932 &\\
\cline{1-4}
103050 &{\color{blue} \href{https://supreme.justia.com/cases/federal/us/304/458/}{Johnson v. Zerbst}} &304 U.S. 458 &1938 &\\
\cline{1-4}
105547 &{\color{blue} \href{https://supreme.justia.com/cases/federal/us/354/476/}{Roth v. United States}} &354 U.S. 476 &1957 &\\
\cline{1-4}
\end{tabular}
}
\end{center}
\end{table*}
\subsubsection{Legal Citation Networks}
Given a citation network of documents, for example, scientific papers or legal opinions, we want to identify similar documents. These could be groundbreaking works that serve to open new fields of discourse in science and law, or foundational works that span disciplines but have less of an impact on discourse, such as ``methods'' papers in science. 

For the purpose of case study, we collected a large digitized record of federal court opinions from the CourtListener project\footnote{\url{https://www.courtlistener.com/}}. The most cited legal decisions from the United States Supreme Court are selected and ego-networks of citations are constructed for these legal cases. Two distinct graph patterns are observed. One is ``Citations have a few giant hubs" and ``Citations are well connected". We list a few examples in Table~\ref{table:legal}, where the titles of the cases with different citation patterns are colored as red and blue, respectively. The ego-networks of the first six cases listed in Table~\ref{table:legal} have just a few giant hubs which are linked by many other cases. For example, the case ``Buckley v. Valeo, 424 U.S. 1 (1976)" is a landmark decision in American campaign finance law. The case ``Coleman v. Miller, 307 U.S. 433 (1939)" is a landmark decision centered on the Child Labor Amendment, which was proposed for ratification by Congress in 1924. These cases are generally centered on a specific topic, and their citations may have a narrowed topic. There are only a few hubs cited frequently by others and the citations generally do not cite each other. On the other side, the ego-networks of the last six cases listed in Table~\ref{table:legal} have citations that are well connected. For example, ``Yick Wo v. Hopkins, 118 U.S. 356 (1886) " was the first case where the United States Supreme Court ruled that a law that is race-neutral on its face, but is administered in a prejudicial manner; The case ``Stromberg v. California, 283 U.S. 359 (1931)" is a landmark in the history of First Amendment constitutional law to include a protection of the substance of the First Amendment. These cases are influential in the history and cited by many diverse subsequent legal decisions, which usually cite each other.

Our Network Vector algorithm is used to learn two-dimensional embeddings from the ego-networks of the legal cases, and their projections are shown as open dots in the right figure of Table~\ref{table:legal}. The structures of the ego-networks for four sampled Supreme Court legal cases (Case IDs: 110578, 93272,101957,105547) are also illustrated in the figure. Note that the ego network includes the case itself (does not show in the figure), and all the cases it cites (smaller circle on the right) as well as all the cases that cite it (larger circle on the left). Lines represent citations among these cases. The two groups of ego-networks contrast each other. Compared to the ego-networks in red boxes, which is cited by unrelated legal cases, there is clearly more coherence in discourse related to the cases in blue boxes, as indicated by citations among other Supreme Court cases that cite this one. Although the differences between these two ego-networks could be captured in a standard way, by features related to the degree distribution of the ego-networks, or their clustering coefficients, the distinctions between other ego-networks may be more subtle necessitating a new approach for evaluating their similarity. In this representation, the position of the case in the learned space captures the similarity of the structure of their ego-networks. Cases that are more similar to the ego-networks in red boxes fall in the top half of the 2-D plane (red open dots); while cases similar to those in blue boxes fall in the bottom half (blue open dots). Thus, distances between the learned representations of the ego-networks of legal cases can be used to quantitatively capture their similarity. 

\section{Experiments}
Network Vector learns representations of network nodes and the entire network simultaneously. We evaluate both representations 
on predictive tasks. First, we apply Network Vector to a setting where only local information about nodes, such as their immediate neighbors, is available. We learn representations for ego-networks of a few nodes using Network Vector and evaluate on role discovery in social networks and concept analogy in encyclopedia. Second, when the information of node connectivities in the entire network is available, we may learn node representations using Network Vector, where the additional global vector for the network is used to help in learning high-quality node representations. The resulting node representations are evaluated on multi-label classification.
\begin{table*}
\centering
  \caption{
Performance of role discovery task. \label{tb:rolediscovery}}
{\footnotesize
\begin{tabular}{| l | c | c | c | c | c | c | c | c | c |}
\hline
&\multicolumn{3}{c|}{\textbf{Wiki - 15 classes}} &\multicolumn{3}{c|}{\textbf{Email - 9 classes}} &\multicolumn{3}{c|}{\textbf{Email - 3 classes}} \\
\hline
Method & p@1 &p@5 &p@10  & p@1 &p@5 &p@10 &p@1 &p@5 &p@10\\
\hline
Degrees+Clustering+Eigens &0.160 &0.149 &0.146 &0.090 &0.102 &0.083 &0.210 &0.200 &0.196\\
node2vec &0.231 &0.224 &0.218 &\textbf{0.290} &0.280 &0.268 &0.500  &\textbf{0.498} &0.474\\
Network Vector &\textbf{0.607} &\textbf{0.560} &\textbf{0.522}  &\textbf{0.290}  &\textbf{0.298}  &\textbf{0.281} &\textbf{0.520}  &\textbf{0.498} &\textbf{0.483} \\
 \hline
   \end{tabular}
}
\end{table*}
\subsection{Role Discovery}
Roles reflect individuals' functions within social networks.
For example, email communication network within an enterprise reflects employees' responsibilities and organizational hierarchies. An engineer's interactions with her team are different from those of a senior manager's. In the Wikipedia network, each article cites other concepts that explain the meaning of the article's concept. Some concepts may ``bridge" the network by connecting different concept categories. For example, the concept \textit{Bat} belongs to the category \textit{Mammals}, however since a bat resembles a bird, it refers to many similar articles about the category \textit{Birds}.

\subsubsection{Datasets}
We use the following datasets in the evaluation:
\begin{itemize}
\item{Enron Email Network:} It contains email interaction data from about 150 users, mostly senior management of Enron. There are about half million emails communicated by 85,601 distinct email addresses\footnote{\url{http://www.cs.cmu.edu/~enron/}}. 
We have 362,467 links left after removing duplicates and self links. Each of the email addresses belonging to Enron employees has one of 9 different positions: \textit{CEO}, \textit{President}, \textit{Vice President}, \textit{Director}, \textit{Managing Director}, \textit{Manager}, \textit{Employee}, \textit{In House Lawyer} and \textit{Trader}. We use the positions as roles. This categorization is fine-grained. In order to understand how the feature representations can reflect the properties of different stratum in the corporation, we also use coarse-grained labels \textit{Leader} (aggregates \textit{CEO}, \textit{President}, \textit{Vice President}), \textit{Manager} (aggregates \textit{Director}, \textit{Managing Director}, \textit{Manager}) and \textit{Employee} (includes  \textit{Employee}, \textit{In House Lawyer} and \textit{Trader}) to divide the users into 3 roles.
    
\item{Wikipedia for Schools Network:} We use a subset of articles available at Wikipedia for Schools\footnote{\url{http://schools-wikipedia.org/}}. This dataset
contains 4,604 articles 
and 119,882 links between them. The articles are categorized by subjects. For example, the article about \textit{Cat} is categorized as \textit{subject.Science.Biology.Mammals}. We use one of 15 second-level category names (e.g., \textit{Science} in the case of  \textit{Cat}) as the role label. 
\end{itemize}



\subsubsection{Methods for Comparison}
For real-world networks, such as email, information about all connectivities of nodes may not be fully available, e.g., for privacy reasons.
For this reason, we explore prediction task with local information (i.e., immediate neighbors). For each node, we first generate its ego-network, which represents the induced subgraph of its immediate neighbors, and learn global vector representations for the set of ego-networks through Network Vector. We use the architecture as in Eq. (\ref{eq:objective}). In our experiments, we repeat $\gamma=10$ times for root node initialization in random walks and the length of each random walks is fixed as $l=80$. For comparison, we evaluate the performance of Network Vector against the following network feature-based algorithms~\cite{berlingerio2012netsimile}: 

\begin{itemize}
\item{Degrees}: 
number of nodes and edges, average node degree, maximum ``in" and ``out" node degrees. The degree features are aggregated to form the representations of the ego-networks.
\item{Clustering Coefficients}: measure the degree to which nodes tend to cluster. 
    We compute global clustering coefficient and average clustering coefficient of nodes for representing each ego-network.
\item{Eigens}: For each ego-network, we compute 10 largest eigenvalues of its adjacency matrix. 

\item{node2vec}~\cite{grover2016node2vec}: This approach learns low-dimensional feature representations of nodes in a network by interpolating between BFS and DFS for sampling node sequences. A parameter $p$ and $q$ is introduced to control the likelihood of revisiting a node in walks, and to dis/encourage outward exploration, resulting in BFS/DFS like sampling strategy. It's interesting to note when $p=1$ and $q=1$, node2vec boils down to DeepWalk~\cite{perozzi2014deepwalk}, which utilizes uniform random walks. We adapt node2vec, and use the mean of learned node vectors to represent each ego-network.
\end{itemize}

\subsubsection{Results}
Given a node's ego-network, we rank other nodes' ego-networks by their distance to it in vector space of feature representations.
Table~\ref{tb:rolediscovery} shows the average precision of retrieved nodes with the same roles (class labels) at cut-off $k=1, 5, 10$. For simplicity, Cosine similarity is used to compute the distance between two nodes. From the result, we can see how the global context allows Network Vector outperform node2vec in role discovery.  However, the performance gain is dependent on different datasets. We observe Network Vector performs slightly better than node2vec on Enron email interaction network, while the improvement of performance is over 150\% on Wikipedia network. Compared to the combination of Degrees, Clustering Coefficients and Eigenvalues, the improvement of the two learning algorithms Network Vector and node2vec are outstanding, with over 100\% performance gain in all cases.

\subsection{Concept Analogy}
We also evaluate the feature representations of ego-networks on the analogy task. For Wikipedia network, we follow the word analogy task defined in~\cite{mikolov2013efficient}. Given a pair of Wikipedia articles describing two concepts $(a,b)$, and an article describing another concept $c$. The task aims to find a concept $d$ such that $a$ is to $b$ as $c$ is to $d$. For example, \textit{Europe} is to \textit{euro} as \textit{USA} is to \textit{dollar}. This analogy task can be solved by finding the concept that is closest to  ${\bf{v}}_b - {\bf{v}}_a + {\bf{v}}_c$ in vector space, where the distance is computed using Cosine similarity.

There are 1,632 semantic tuples in Wikipedia network matched for the semantic pairs in~\cite{mikolov2013efficient}. We use them as evaluation benchmark. Table~\ref{tb:analogy} shows the accuracy of hitting the answer $d$ within cut-off $k=1, 5, 10$ positions in the ranking list. From the results, we can see Network Vector performs much better than the baseline, which use degree, clustering coefficients and eigenvalues of the adjacency matrix. The combination of heterogeneous features (degrees, clustering coefficients and eigenvalues) in different scale causes the difficulty to utilize an efficient distance metric. However, Network Vector does not suffer from this problem by automating the feature learning using an objective function. In this task, we empirically fix the dimensionality of vectors as 100 and context window as 10.
\begin{table}
\centering
  \caption{
Performance of concept analogy task. \label{tb:analogy}}
{\footnotesize
\begin{tabular}{| l | c | c | c | }
\hline
\textbf{Method} & \textbf{Hit@1} &\textbf{Hit@5} &\textbf{Hit@10}\\
\hline
Degrees &0.0147 &0.0423 &0.0717\\
Clustering Coefficients &0.0006 &0.0043 &0.0086 \\
Eigens &0.0025 &0.0074 &0.0116 \\
Degrees+Clustering+Eigens &0.0153 &0.0453 &0.0803 \\
node2vec &0.2450 &0.5098 &0.6150 \\
Network Vector &\textbf{0.2849}  &\textbf{0.5619}  &\textbf{0.6930}  \\
 \hline
   \end{tabular}
}
\end{table}
\subsection{Multi-label Classification}
Multi-label classification is a challenge task, where each node may have one or multiple labels. A classifier is trained to predict multiple possible labels for each test node.
In our Network Vector algorithm, the global representation of entire network serves as additional context along with local neighborhood in learning node representations. 

\subsubsection{Datasets}
To understand whether the global representation helps learning better node representation, we perform multi-label classification with the same benchmarks and experimental procedure as~\cite{grover2016node2vec} using the same datasets:
\begin{itemize}
\item{BlogCatalog~\cite{zafarani2009social,tang2009relational}}: This is a network of social relationships provided by bloggers on the BlogCatalog website. The labels represent the interests of bloggers on a list of topic categories. There are 10,312 nodes, 333,983 edges in the network and 39 distinct labels for nodes.
\item{Protein-Protein Interactions (PPI)~\cite{breitkreutz2008biogrid,grover2016node2vec}}: This is a subgraph of the entire PPI network for Homo Sapiens. The node labels are obtained from hallmark gene sets~\cite{liberzon2011molecular} and represent biological states. There are 3,890 nodes, 76,584 edges in the network and 50 distinct labels for nodes.
\item{Wikipedia Cooccurrences~\cite{mahoney2009large,grover2016node2vec}}: This is a network of words appearing in the first million bytes of the Wikipedia dump. The edge weight is defined by the cooccurrence of two words within a 2-length slide window. The Part-of-Speech (POS) tags~\cite{marcus1993building} inferred using the Stanford POS-Tagger~\cite{toutanova2003feature} are used as labels. There are 4,777 nodes, 184,812 edges in the network and 40 distinct labels for nodes.
\end{itemize}

\subsubsection{Methods for Comparison}
We compare the node representations learned by Network Vector against the following feature learning methods for node representations:
\begin{itemize}
\item{Spectral clustering}~\cite{tang2011leveraging}: This method learns the $d$-smallest eigenvectors of the normalized graph Laplacian matrix, and utilize them as the $d$-dimensional feature representations for nodes.
\item{DeepWalk}~\cite{perozzi2014deepwalk}: This method learns $d$-dimensional feature representations using Skip-gram~\cite{mikolov2013efficient,mikolov2013distributed} from node sequences, that are generated by uniform random walks from the source nodes on graph.
\item{LINE}~\cite{tang2015line}: This method learns $d$-dimensional feature representations by sampling nodes at 1-hop and 2-hop distance from the source nodes in BFS-like manner.
\item{node2vec}~\cite{grover2016node2vec}: We use the original node2vec algorithm with optimal parameter settings of $(p, q)$ reported in ~\cite{grover2016node2vec}.
\end{itemize}

Network Vector utilizes only first-order or second-order proximity between nodes in two-layer neural embedding framework. The first layer computes the context feature vector, and the second layer computes the probability distribution of target nodes. It is similar to other neural embedding based feature learning methods DeepWalk, LINE and node2vec. For fair comparison, we exclude recent approaches GraRep~\cite{cao2015grarep}, HNE~\cite{chang2015heterogeneous} and SDNE~\cite{wang2016structural}. It is because GraRep utilizes information from network neighborhoods beyond second-order proximity, and both HNE and SDNE employ deep neural networks that have multiple layers (more than two). GraRep, HNE and SDNE are less computational efficient and cannot scale well, as compared to DeepWalk, LINE, node2vec and our algorithm Network Vector.
\begin{table}
\centering
  \caption{
Macro-F1 scores for multi-label classification with a balanced $50-50$ split between training and testing data. Results of Spectral clustering, DeepWalk, LINE and node2vec are reported in node2vec paper. \label{tb:multi-label-classification} }
{\footnotesize
\begin{tabular}{| l | c | c | c |}
\hline
\textbf{Algorithm} &\multicolumn{3}{c|}{\textbf{Dataset}}  \\
\cline{2-4}
&BlogCatalog &PPI &Wikipedia \\
 \hline
Spectral Clustering &0.0405 &0.0681  &0.0395 \\
LINE &0.0784  &0.1447 &0.1164 \\
DeepWalk &0.2110 &0.1768 &0.1274 \\
node2vec (p*, q*) &0.2581  &0.1791 &0.1552 \\
\hline
Network Vector (p=q=1)&0.2473 &0.1938 &0.1388 \\
Network Vector (p*, q*)&\textbf{0.2607} &\textbf{0.1985} &\textbf{0.1765} \\
\hline
settings (p*, q*) &0.25, 0.25 &4, 1 &4, 0.5 \\
\hline
Gain over DeepWalk &12.4\% &9.6\% &8.9\% \\
Gain over ndoe2vec &1.0\% &9.7\% &13.7\% \\
\hline
   \end{tabular}
}
\end{table}
For fair comparison, we use the inverse architecture of Network Vector, which is Skip-gram~\cite{mikolov2013efficient} like and similar to that of node2vec. The parameter settings for Network Vector are in favor of node2vec, and exactly the same as in~\cite{grover2016node2vec}. Specifically, we set $d = 128$, $r = 10$, $l = 80$, and a context size $n = 10$, and are aligned with typical values used for DeepWalk and LINE. A single pass of the data (one epoch) is used for optimization. In order to perform multi-label classification, the learned node representations from each approach are used as feature input to a one-vs-rest logistic regression with L2 regularization. Our experiments are repeated for 10 random equal splits of train and test data, and average results are reported. 

\subsubsection{Results}
Macro-F1 scores are used as evaluation metrics, and Table~\ref{tb:multi-label-classification} shows the results. We run Network Vector with node sequences generated by biased random walks from node2vec. The default parameter setting $(p=1, q=1)$ used in DeepWalk and the optimal parameter setting of node2vec reported in~\cite{grover2016node2vec} are used. 

From the results, we can see Network Vector outperforms node2vec using the same biased random walks, and DeepWalk using the same uniform random walks. It is evident that the global representation of the entire network allows Network Vector to exploit the global structure of the networks to learn better node representations. Network Vector achieves a slight performance gain, $1.0\%$ over node2vec, and a significant $12.4\%$ gain over DeepWalk on BlogCatalog. As we can see on PPI, The gain of Network Vector over node2vec and DeepWalk are significant and similar, $9.6\%$ and $9.7\%$ respectively. In the case of Wikipedia word cooccurrence network, Network Vector outperforms node2vec with a decent margin, achieving $13.7\%$ performance gain, while with a less gain, $8.9\%$ over DeepWalk. Overall, sampling strategies even with optimal parameter settings $(p, q)$ in node2vec are limited in exploration of local neighborhood of the source nodes, but cannot exploit the global network structure well. Network Vector overcomes the limitation of locality. By utilizing an additional global vector to memorize the collective information from all the local neighborhoods of nodes even within 2-hops, Network Vector learns improved node representations.
\begin{figure}
\centering
\begin{tabular}{@{}c@{}c}
   \includegraphics[width=0.25\textwidth]{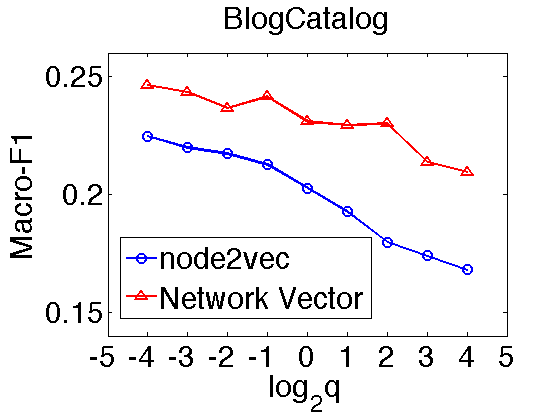} &
   \includegraphics[width=0.25\textwidth]{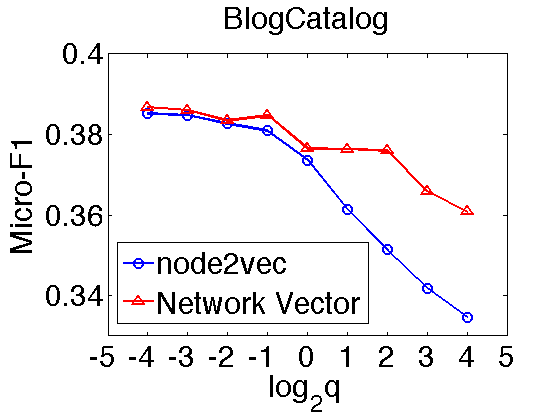} \\
   \includegraphics[width=0.25\textwidth]{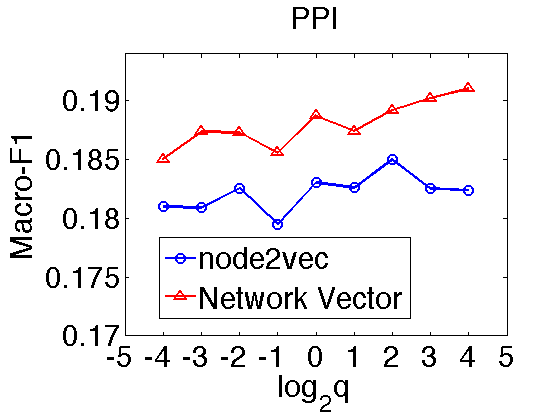} &
   \includegraphics[width=0.25\textwidth]{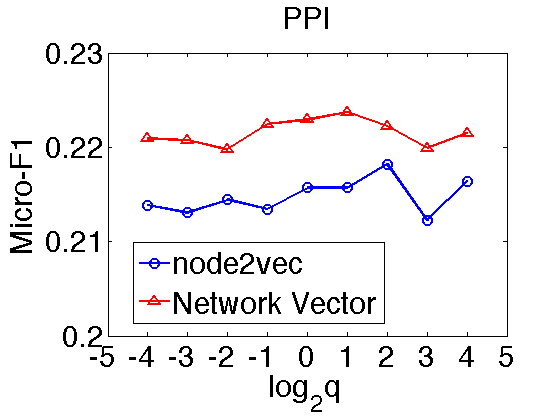} \\
   \includegraphics[width=0.25\textwidth]{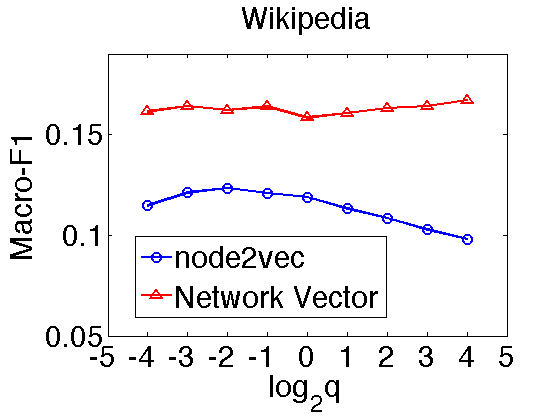} &
   \includegraphics[width=0.25\textwidth]{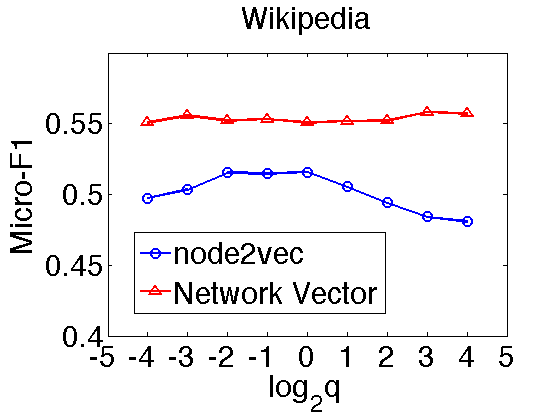} \\
\end{tabular}
   \caption{Performance of Network Vector and node2vec on varying the parameter $q$ when fixing $p=\infty$ to encourage reaching unvisited nodes in random walks. Macro-F1 and Micro-F1 scores for multi-label classification with a balanced $50\%$ train-test split are reported. \label{fig:q}}
 \end{figure}

\subsubsection{Parameter Sensitivity}
In order to understand how Network Vector improves in learning node representations with biased random walks in fine-grained settings, we evaluate performance while varying the parameter settings of $(p, q)$. We fix $p=\infty$ to discourage revisiting sampled nodes at the previous step in random walks, and varying the value $q$ in the range from $2^{-4}$ to $2^4$ to perform DFS-like sampling in various degrees. 

Figure~\ref{fig:q} shows the comparison results for Network Vector and ndoe2vec in both Macro-F1 and Micro-F1 scores. As we can see, Network Vector consistently outperforms node2vec in different parameter settings of $q$ in all the three datasets. However, we observe on BlogCatalog, Network Vector achieves relatively larger gains over node2vec when $q$ is large that the random walks is biased towards BFS-like sampling, as compared to that when $q$ is small that the sampling is more DFS-like. It is mainly because when the random walks is biased towards nodes close to the source nodes, the global information of network structure that are exploited by Network Vector can compensate more for locality information using BFS-like sampling. However, when $q$ is small, the random walks is biased towards sampling nodes far away from the source nodes, and explore information close to the global network structure. Hence, Network Vector is not quite helpful in this case. We can see similar patterns of performance margin between Network Vector and node2vec when $q$ tends to be large in word cooccurrence network of Wikipedia. However, in the case of PPI, the performance gains achieved by Network Vector over node2vec are stable even various values of $q$ are used. The reason is probably because the biological states of proteins in a protein-protein interaction network exhibit a high degree of homophily, since proteins in local neighborhood usually organize together to perform similar functions. Hence, the global network structure is not quite informative to predict the biological states of proteins as we set a large value of $q$.
\begin{figure}
\centering
   \begin{tabular}{@{}c@{}c}
   \includegraphics[width=0.25\textwidth]{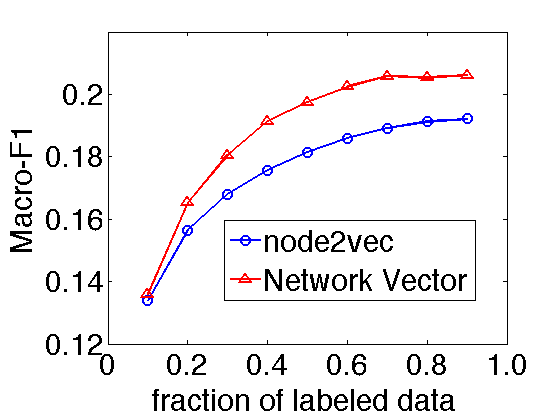} &
   \includegraphics[width=0.25\textwidth]{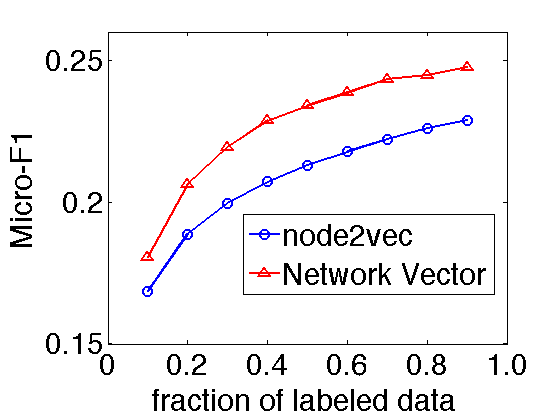} \\
\end{tabular}
   \caption{Performance of Network Vector and node2vec on varying the faction of labeled data for training. \label{fig:labeleddata} }
   \end{figure} 

\subsubsection{Effect of Training Data}
To see the effect of training data, we compare performance while varying the fraction of labeled data from $10\%$ to $90\%$. Figure~\ref{fig:labeleddata} shows the results on PPI. The parameters $(p, q)$ is fixed using optimal values $(4, 1)$. As we can see, when using more labeled data, the performance of node2vec and Network Vector generally increases. Network Vector achieves the largest gain over node2vec of $9.0\%$ in Macro-F1 score and $10.3\%$ at $40\%$ labeled data. When only $10\%$ labeled data is used,  Network Vector only yields $1.5\%$ gain in Macro-F1 score, and $7.1\%$ in Micro-F1 score. We have similar observations on BlogCatalog and Wikipedia datasets, and the results are not shown.
\section{Conclusion}
We have presented Network Vector, an algorithm for learning distributed representations of nodes and networks simultaneously. By embedding the network in a lower-dimensional vector space, our algorithm allows for quantitative comparison of networks. It also allows for the comparison of individual network nodes, since each node can be represented by its ego-network---a network containing the node itself, its network neighbors, and all connections between them. 

In contrast to existing network embedding methods, which only learn representations of component nodes, Network Vector directly learns the representation of an entire network. Learning a representation of a network allows us to evaluate the similarity between two networks or two individual nodes, which enables us to answer questions that were difficult to address with existing methods. For instance, given a node in a network, for example, a manager within an organization, we can identify other people serving a similar role within that organization. Also, given a connection, denoting some relationship between two people within a social network, we could find another pair in an analogous relationship. Beyond social networks, we can also answer new questions about knowledge networks that connect concepts or documents to each others, for example, Wikipedia and citations networks. This can be useful especially in cases where the contents of documents is not available for privacy or other reasons, but the network of interactions exists.

For the networks in which content is available for the nodes, the learning method could be extended to account for it. For example, for knowledge networks, the approach could be combined with text to learn representations of networks that will give a more fine-grained view of their similarity. Additionally, other non-textual attributes could also be included in the learning algorithm. The flexibility of such learning algorithms make them ideal candidates for applications requiring similarity comparison of different types of objects.

\bibliographystyle{named}
\bibliography{hao}
\flushend

\end{document}